\documentclass[11pt,english,final,aps,prc,tightenlines,superscriptaddress]{revtex4}
\usepackage[T1]{fontenc}
\usepackage[latin9]{inputenc}
\usepackage{float}
\usepackage{graphicx}

\makeatletter

\usepackage{float}

\usepackage{esint}



\makeatother

\usepackage{babel}

\begin{document}

\title{Computational Model for Electron-Nucleon Scattering and Weak Charge
of the Nucleon}

\author{A. Aleksejevs}

\affiliation{Division of Science of SWGC, Memorial University, Corner Brook, NL,
Canada}

\author{S. Barkanova}

\affiliation{Physics Department, Acadia University, Wolfville, NS, Canada}

\author{P. G. Blunden}

\affiliation{Department of Physics and Astronomy, University of Manitoba, Winnipeg,
MB, Canada}

\date{\today}

\begin{abstract}
We show how computational symbolic packages such as \emph{FeynArts,
FormCalc, Form }and \emph{LoopTools} can be adopted for the evaluation
of one-loop hadronic electroweak radiative corrections for electron-nucleon
scattering and applied to calculations of the nucleon weak charge.
Several numerical results are listed, and found to be in good agreement
with the current experimental data. 
\end{abstract}
\maketitle

\section{Introduction}

Next-to-Leading-Order (NLO) effects in electroweak interactions play
a crucial role in tests of the Standard Model, and require careful
theoretical evaluation. An excellent place to search for new physics,
the deviation of the weak charge of the nucleon from its Standard
Model prediction also requires considerable experimental and theoretical
input. The importance of theoretical predictions for the weak charge
of the nucleon was recognized more that two decades ago by Marciano
and Sirlin \citep{MS84}. Their original analysis was followed recently
by significant theoretical work on the proton weak charge \citep{EKM2003},
which presented an updated Standard Model prediction for the weak
charge of the proton, and estimated the QCD structure uncertainties.

Computer packages such as \emph{FeynArts} \citep{FeynArts}, \emph{FormCalc}
\citep{FormCalc}, \emph{LoopTools }\citep{LoopTools}, \citep{FF}
and \emph{Form }\citep{Form} have created the possibility to go one
step further. Now, we have an option of calculating parity-violating
NLO effects including all of the possible loop contributions within
a given model. We have already tested the automated approach to calculate
one-quark radiative corrections \citep{BAB2002}.

In the work presented here, we adopt \emph{FeynArts} and \emph{FormCalc}
for the NLO symbolic calculations of amplitude or differential cross
section in electron-nucleon scattering. Using Dirac and Pauli-type
couplings with the monopole form factor approximation, we construct
the computational model enabling \emph{FeynArts} and \emph{FormCalc}
to deal with electron-nucleon scattering up to the NLO level. Since
the weak charge of the nucleon is directly related to the form factors
of the parity-violating part of the amplitude at the zero momentum
transfer, we choose the calculation of the weak charge of the nucleon
to be a suitable test our method. Within the uncertainty, which mostly
comes from the uncertainty of the current electromagnetic form factor
measurements, our results agree with experiment.

We start with expressions for Pauli and Dirac couplings in terms of
fermion weak and electric charges, and the definition and classification
of one-loop radiative corrections. After that, an example for the
$\left\{ \gamma-Z\right\} $ box type of correction for electron-proton
scattering is considered and a computational model is proposed. We
proceed with computational details for the self-energy graphs and
vertex correction graphs. Since we reserve full kinematic dependence
in all types of our NLO calculations, it will make it easier to adopt
our results to the current and future parity-violating experiments.

\section{Theory}

\subsection{Dirac and Pauli Coupling}

In the approximation where the nucleon behaves as a point-like particle,
vector boson couplings obey general rules of electroweak theory. For
left and right handed fermions, we can use the following structure
for the $\left\{ Z-N\right\} $ type couplings: \begin{equation}
\Gamma_{\mu}^{Z-N}=ie\,\left[g_{L}^{Z-N}\gamma_{\mu}\varpi_{-}+g_{R}^{Z-N}\gamma_{\mu}\varpi_{+}\right],\label{e1.1}\end{equation}
 where $\varpi_{\pm}=\frac{1\pm\gamma_{5}}{2}$ are chirality projectors
and $g_{L,R}^{Z-N}$ represents the coupling strength for the left
and right handed fermions, respectively. Substitution of $\varpi_{\pm}$
into Eq.(\ref{e1.1}) yields the vector and axial vector representation
of the coupling $\Gamma_{Z-N}^{\mu}$: \begin{equation}
\Gamma_{\mu}^{Z-N}=ie\,\left[\frac{g_{L}^{Z-N}+g_{R}^{Z-N}}{2}\gamma_{\mu}+\frac{g_{R}^{Z-N}-g_{L}^{Z-N}}{2}\gamma_{\mu}\gamma_{5}\right].\label{e1.2}\end{equation}
 Couplings of vector bosons to fermions derived from the neutral current
part of the electroweak Lagrangian predict tree-level values of $g_{L}^{Z-N}$
and $g_{R}^{Z-N}$ given by: \begin{eqnarray}
\frac{g_{L}^{Z-N}+g_{R}^{Z-N}}{2} & = & \frac{1}{4}\frac{C_{3}-4s_{w}^{2}Q}{c\,_{w}s_{w}},\nonumber \\
\label{e1.3}\\\frac{g_{R}^{Z-N}-g_{L}^{Z-N}}{2} & = & -\frac{1}{4}\frac{C_{3}}{c_{w}\, s_{w}}.\nonumber \end{eqnarray}
 Here, $C_{3}$ and $Q$ are twice the value of the isospin and electric
charge respectively, and $s_{w}$ and $c_{w}$ are $\sin\theta_{W}$
and $\cos\theta_{W}$, where $\theta_{W}$is a Weinberg mixing angle.
In the case when a photon couples to the nucleon, $g_{L}^{\gamma-N}=g_{R}^{\gamma-N}=Q$
and $\Gamma_{\mu}^{\gamma-N}=ieQ\gamma_{\mu}$.

To accommodate nucleon structure, we have the couplings preserve their
vector and vector-axial structure, but with the charges replaced by
the corresponding form factors. The electromagnetic $\Gamma_{\mu}^{\gamma-N}$
coupling has two vector components responsible for static electric
and magnetic interactions: \begin{equation}
\Gamma_{\mu}^{\gamma-N}\left(q\right)=ie\left[F_{1}\left(q\right)\gamma_{\mu}+\frac{i}{2m_{N}}\sigma_{\mu\alpha}q^{\alpha}F_{2}\left(q\right)\right],\label{e1.5}\end{equation}
 where $F_{1}\left(q\right)$ and $F_{2}\left(q\right)$ are the Dirac
and Pauli form factors, respectively, and $q^{\alpha}$ is the four-momentum
transferred to the nucleon. As for $\Gamma_{\mu}^{Z-N}\left(q\right)$,
we have: \begin{equation}
\Gamma_{\mu}^{Z-N}\left(q\right)=ie\left[f_{1}\left(q\right)\gamma_{\mu}+\frac{i}{2m_{N}}\sigma_{\mu\alpha}q^{\alpha}f_{2}\left(q\right)+g_{1}\left(q\right)\gamma_{\mu}\gamma_{5}\right],\label{e1.6}\end{equation}
 with $f_{1}\left(q\right)$, $f_{2}\left(q\right)$ and $g_{1}\left(q\right)$
as weak Dirac, Pauli, and axial-vector form factors. According to
the first line of the Eq.(\ref{e1.3}), form factors $f_{1}\left(q\right)$
and $f_{2}\left(q\right)$ are expressed as: \begin{equation}
f_{1,2}\left(q\right)=\frac{1}{4c_{w}s_{w}}\left(F_{1,2}^{V\left(N\right)}\left(q\right)-4s_{w}^{2}F_{1,2}\left(q\right)\right),\label{e1.7}\end{equation}
 with $F_{1,2}^{V\left(p\right)}=-F_{1,2}^{V\left(n\right)}=F_{1,2}^{p}-F_{1,2}^{n}$.
For $g_{1}\left(q\right)$ we have \begin{equation}
g_{1}\left(q\right)=-\frac{1}{4c_{w}s_{w}}g_{A}\left(q\right),\label{e1.8}\end{equation}
 where $g_{A}^{p}\left(q\right)=-g_{A}^{n}\left(q\right)=g_{A}\left(q\right)$
is an axial form factor. To considerably simplify analytical expressions,
in our computational model we use the monopole structure for form
factors expressed as \begin{equation}
\left\{ F_{1,2},g_{A}\right\} \left(q\right)=\frac{\Lambda^{2}\left\{ F_{1,2},g_{A}\right\} \left(0\right)}{\Lambda^{2}-q^{2}},\label{e1.9}\end{equation}
 which is a reasonable approximation in our case. The value of the
parameter $\Lambda^{2}=0.83m_{N}^{2}$ is found after the fit of the
electromagnetic form factors by monopole approximation in the low
momentum transfer region. Further details on how couplings where implemented
in the computational model are given in the appendix.

\subsection{Definition of NLO Contribution}

In order to define the Next-to-Leading-Order hadronic corrections,
we are using the electron-nucleon parity violating Hamiltonian in
the form proposed by \citep{MS84}: \begin{equation}
H^{PV}=\frac{G_{F}}{\sqrt{2}}\left[C_{1N}\left(\overline{u}_{e}\gamma^{\mu}\gamma_{5}u_{e}\right)\left(\overline{u}_{N}\gamma_{\mu}u_{N}\right)+C_{2N}\left(\overline{u}_{e}\gamma^{\mu}u_{e}\right)\left(\overline{u}_{N}\gamma_{\mu}\gamma_{5}u_{N}\right)\right].\label{i1.4}\end{equation}
 Form factors $C_{1N}$ and $C_{2N}$ represent perturbative expansion
resulting in \begin{equation}
C_{\left\{ 1,2\right\} N}=\sum_{i}C_{\left\{ 1,2\right\} N}^{i}=C_{\left\{ 1,2\right\} N}^{0}+C_{\left\{ 1,2\right\} N}^{1}+O\left(\alpha^{3}\right).\label{i1.5}\end{equation}
 The superscript in $C_{\left\{ 1,2\right\} N}^{i}$ represents the
order of the perturbation ({}``zero''- tree level, {}``one'' -
one loop level (NLO) and so on). Here $C_{\left\{ 1,2\right\} N}^{1}$
can be defined as an one-loop contribution to the parity-violating
form factor, normalized by the Fermi constant, $G_{F}=\frac{\alpha\pi}{\sqrt{2}m_{W}^{2}s_{W}^{2}}$.
A calculation of NLO form factors is related to the calculations of
loop integrals represented by three topological classes: box, self
energy, and vertex (triangle) graphs. To preserve gauge invariance
we have to include all the possible bosons of the Standard Model in
these topological classes. Taking into account that in the t'Hooft-Feynman
gauge the contribution coming from the Higgs scalar and gauge fixing
fields is negligible, we choose to consider boxes and nucleon vertex
corrections (triangles) with $\gamma,\, Z$ and $W^{\pm}$ vector
bosons only. For the rest of the graphs -- self-energies and lepton
vertex corrections -- we accounted for all the possible particles
in the Standard Model. Accordingly, we consider NLO corrections for
every class.

\subsection{Box Diagrams in $e-N$ Scattering}

The precise formulae for an entire set of graphs are cumbersome, and
it is not feasible to show them in the present work. Here, we provide
some details for $\left\{ \gamma-Z\right\} $ boxes only. Our complete
calculations are shown in \citep{ThesisAA}. The full set of diagrams
applicable for this case can be found in \citep{ThesisSB}.

According to the Feynman rules (see, for example \citep{BHS86}),
the amplitude for a $\left\{ \gamma-Z\right\} $ box (see Fig.(\ref{f1.1}))
can be written as:%
\begin{figure}
\begin{centering}
\includegraphics{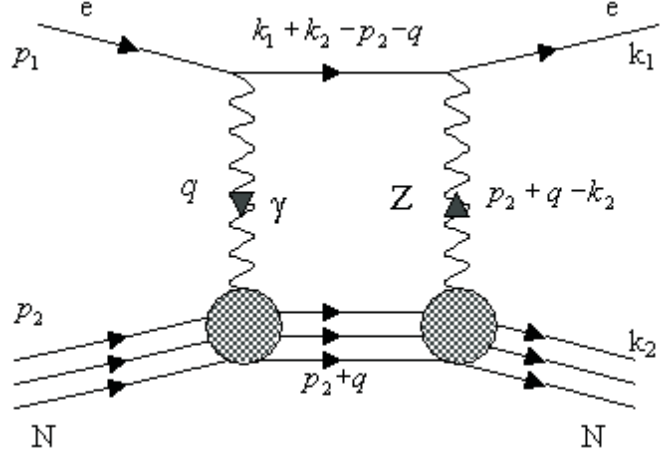} 
\par\end{centering}

\caption{$\gamma-Z$ box diagram in the electron-nucleon scattering. \label{f1.1}}

\end{figure}

\begin{eqnarray}
M^{\left\{ \gamma-Z\right\} } & = & \frac{1}{16\pi^{4}}\int d^{4}q\left(i\overline{u}_{e}\Gamma_{\mu}^{Z-e}\frac{m_{e}+\not k_{1}+\not k_{2}-\not q-\not p_{2}}{\left(p_{2}-k_{1}-k_{2}+q\right)^{2}-m_{e}^{2}}\Gamma_{\nu}^{\gamma-e}u_{e}\right)\nonumber \\
\label{e1.4}\\ &  & \ \ \left(i\overline{u}_{N}\Gamma_{\rho}^{Z-N}\frac{m_{N}+\not q+\not p_{2}}{\left(p_{2}+q\right)^{2}-m_{N}^{2}}\Gamma_{\alpha}^{\gamma-N}u_{N}\right)\left(\frac{g^{\nu\alpha}}{q^{2}}\frac{g^{\mu\rho}}{\left(p_{2}+q-k_{2}\right)^{2}-m_{Z}^{2}}\right).\nonumber \end{eqnarray}

\begin{figure}
\begin{centering}
\includegraphics[scale=0.4]{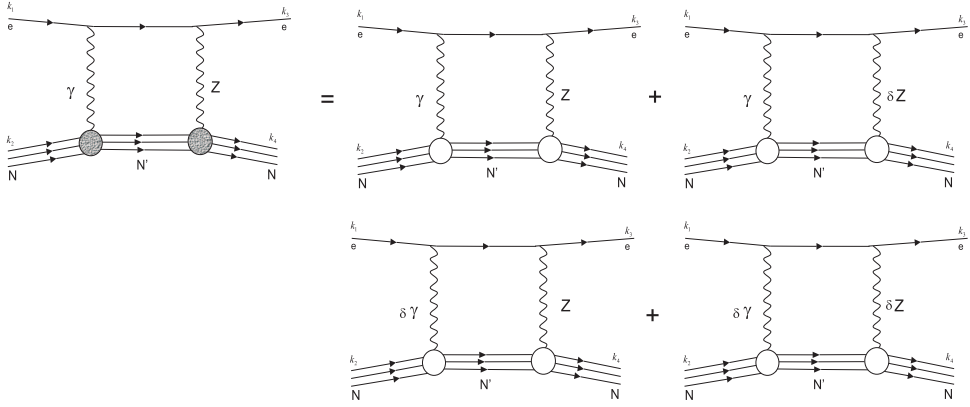} 
\par\end{centering}

\caption{Expansion of the $\gamma-Z$ box in terms of $\delta\gamma,\,\delta Z$
particles\label{fhdr1.3}}

\end{figure}

In order to integrate the amplitude in Eq.(\ref{e1.4}) and the rest
of the NLO loop integrals using \emph{FeynArts} and \emph{FormCalc},
it is necessary first to create a model file with declarations of
all the fields participating in interaction and second to describe
all the possible couplings of the fields on both generic and particle
levels. For the case of electron-nucleon scattering, the only new
particles in the model besides the particles of the Standard Model
are neutron and proton, so it is rather straightforward to declare
them in the \emph{FeynArts} model file. On other hand, the description
of the couplings according to the Eq.(\ref{e1.5}) and Eq.(\ref{e1.6})
in the \emph{FeynArts} represents a certain problem. Because we use
the monopole form factor approximation (see Eq.(\ref{e1.9})), it
introduces momentum dependence in the coupling's denominator. For
\emph{FeynArts} to deal with momentum-dependent couplings with momentum
dependence introduced in the denominator, these couplings have to
be described as propagators. Of course, a propagator introduced in
the coupling conflicts with the declaration of the fields in the \emph{FeynArts}
model files which use the same propagator notation. We can solve the
problem by transferring the monopole form factor from the coupling
part of model to the part where fields are described using propagator
notation. For boxes that can be achieved by starting with the general
definition of a four-point tensor integral of rank $k$ in the form
\begin{eqnarray}
T_{\mu_{1}...\mu_{k}}^{4} & = & \frac{1}{i\pi^{2}}\int d^{4}q\frac{q_{\mu_{1}}...q_{\mu_{k}}}{\left(\left(p_{2}-k_{1}-k_{2}+q\right)^{2}-m_{e}^{2}\right)\left(\left(p_{2}+q\right)^{2}-m_{N}^{2}\right)}\cdot\nonumber \\
\label{e1.19}\\ &  & \ \ \ \cdot\frac{1}{q^{2}}\cdot\frac{1}{\left(p_{2}+q-k_{2}\right)^{2}-m_{Z}^{2}}.\nonumber \end{eqnarray}
 By adding the monopole form factor approximation into the above definition,
we obtain a six-point tensor integral which can be reduced into a
combination of four-point integrals by using the following expansion:
\begin{eqnarray}
\  &  & \frac{1}{D_{1}D_{2}}=\frac{1}{q^{2}}\cdot\frac{1}{\left(p_{2}+q-k_{2}\right)^{2}-m_{Z}^{2}}\cdot\frac{\Lambda^{4}}{\left(q^{2}-\Lambda^{2}\right)\left(\left(p_{2}+q-k_{2}\right)^{2}-\Lambda^{2}\right)}=\nonumber \\
\label{e1.20}\\ &  & \ \ \ \frac{\Lambda^{2}}{\left(\Lambda^{2}-m_{Z}^{2}\right)}\left(\frac{1}{q^{2}-\Lambda^{2}}-\frac{1}{q^{2}}\right)\cdot\left(\frac{1}{\left(p_{2}+q-k_{2}\right)^{2}-\Lambda^{2}}-\frac{1}{\left(p_{2}+q-k_{2}\right)^{2}-m_{Z}^{2}}\right).\nonumber \end{eqnarray}
 Here $\frac{1}{D_{1}D_{2}}$ represents the last two terms of the
product in Eq.(\ref{e1.19}) multiplied by the two monopole form factors
of the box diagram. Explicitly, the right-hand side of Eq.(\ref{e1.20})
can be written in the form\begin{eqnarray}
\frac{1}{D_{1}D_{2}}=\frac{\Lambda^{2}}{\left(\Lambda^{2}-m_{Z}^{2}\right)}\frac{1}{q^{2}}\frac{1}{\left(p_{2}+q-k_{2}\right)^{2}-m_{Z}^{2}}+\frac{-\Lambda^{2}}{\left(\Lambda^{2}-m_{Z}^{2}\right)}\frac{1}{q^{2}-\Lambda^{2}}\frac{1}{\left(p_{2}+q-k_{2}\right)^{2}-m_{Z}^{2}}+\nonumber \\
\label{e1.20a}\\\frac{-\Lambda^{2}}{\left(\Lambda^{2}-m_{Z}^{2}\right)}\frac{1}{q^{2}}\frac{1}{\left(p_{2}+q-k_{2}\right)^{2}-\Lambda^{2}}+\frac{\Lambda^{2}}{\left(\Lambda^{2}-m_{Z}^{2}\right)}\frac{1}{q^{2}-\Lambda^{2}}\frac{1}{\left(p_{2}+q-k_{2}\right)^{2}-\Lambda^{2}} &  & .\nonumber \end{eqnarray}
 Substitution of the last equation into amplitude Eq.(\ref{e1.4})
results in an expansion where each term of the sum represents an amplitude
constructed for electron-nucleon scattering with the point-like nucleon.
Also, in this consideration, the couplings between nucleon and vector
bosons are multiplied by the factor $\pm\sqrt{\frac{\Lambda^{2}}{\left(\Lambda^{2}-m_{Z}^{2}\right)}}$,
and the structure of the second, third and fourth terms of Eq.(\ref{e1.20a})
suggests the introduction of fictitious vector boson particles $\left\{ \delta\gamma,\delta Z\right\} $
with fixed masses $m_{\left\{ \delta\gamma,\delta Z\right\} }=\Lambda$.
These bosons have no physical meaning, of course, but they allow us
to remove momentum dependence in the denominator of the coupling.
Now we can model couplings between nucleon and vector fields using
Eq.(\ref{e1.5}) and Eq.(\ref{e1.6}) with a monopole form factor
moved into the definition of propagators for the new vector fields
such as $\left\{ \delta\gamma,\delta Z\right\} $ where $\Pi_{\left\{ \delta\gamma,\delta Z\right\} }^{\mu\nu}=\frac{g^{\mu\nu}}{q^{2}-\Lambda^{2}}$.
A diagrammatic representation of the proposed expansion is given by
the set of Feynman graphs in Fig.(\ref{fhdr1.3}). %

As for the boxes, we have all we need to complete the automated construction
of amplitudes in the \emph{FeynArts} and the calculations in the \textit{FormCalc}.
We should note that Eq.(\ref{e1.4}) is written for the $\left\{ \gamma-Z\right\} $
box diagram only. To have a complete analysis, it is necessary to
consider $\left\{ Z-\gamma\right\} $, $\left\{ \gamma-Z\right\} $
and $\left\{ Z-\gamma\right\} $ crossed boxes as well. $\left\{ Z-Z\right\} $
and $\left\{ W-W\right\} $ box diagrams should be considered, too.
Our calculations include 36 boxes in total.

\subsection{Self-Energy Graphs}

\begin{figure}
\begin{centering}
\includegraphics[scale=0.4]{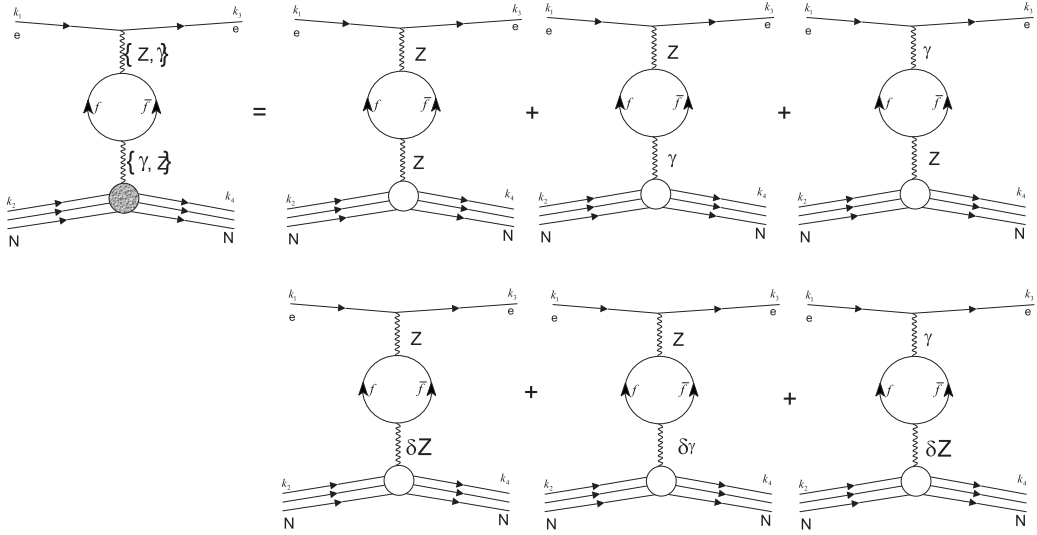} 
\par\end{centering}

\caption{Expansion of the self-energy graphs in terms of $\delta\gamma,\,\delta Z$
partners. Here, factor $\pm\sqrt{\frac{\Lambda^{2}}{\left(\Lambda^{2}-m_{Z}^{2}\right)}}$
used for the couplings of the box diagrams should be replaced by one.\label{fhdr1.4}}

\end{figure}

In total, 116 self-energy graphs and 6 counterterms contribute to
the PV $\left\{ e-N\right\} $ amplitude. This includes gauge and
the gauge-fixing fields, the Higgs field, and virtual leptonic and
quark pairs in creation-annihilation processes in the loops. Moreover,
the vertex $\left\{ N-V-N\right\} $ does not belong to the loop integrals
and plays the role of a multiplicative factor proportional to the
coupling defined in Eq.(\ref{e1.5}) and Eq.(\ref{e1.6}). In \emph{FormCalc}
the self-energy loop integrals can be evaluated using the expansion
given in the Fig.(\ref{fhdr1.4}), and then tensor decomposition and
tensor reduction techniques applied, leaving the final result as a
combination of one and two point scalar integrals which we compute
using Gauss integration subroutines of \emph{LoopTools}. %

To cancel ultraviolet divergences we employ an on-shell renormalization
scheme according to \citep{Hah97}. We have to assume that quarks
in the self-energy loops are free, but it places certain constraints.
Since quarks are confined, a QCD strong quark-quark interaction should
always be considered. It is possible to bypass these complications
by replacing quarks with pions, or use {}``free'' quarks but with
adjusted effective masses. Here, we use the second approach with the
effective mass of the quarks coming from a fit of hadronic vacuum
polarization to the measurements of QED cross section of the process
$e^{+}e^{-}\rightarrow\,$hadrons normalized to the QED $e^{+}e^{-}\rightarrow\mu^{+}\mu^{-}$
cross section. The real part of the renormalized hadronic vacuum polarization
satisfies the dispersion relation: \begin{equation}
\bigtriangleup\alpha_{hadr}^{\gamma}\left(s\right)=-Re\widehat{\,\Pi}_{had}^{\gamma}\left(s\right)=\frac{\alpha}{3\pi}s\int_{4m_{\pi}^{2}}^{\infty}\frac{R^{\gamma}\left(s^{\prime}\right)}{s^{\prime}\left(s^{\prime}-s\right)}ds^{\prime},\label{e1.31}\end{equation}
 with \begin{equation}
R^{\gamma}\left(s\right)=\frac{\sigma\left(e^{+}e^{-}\rightarrow hadrons\right)}{\sigma\left(e^{+}e^{-}\rightarrow\mu^{+}\mu^{-}\right)}\label{e1.32}\end{equation}
 being a very well known experimental quantity and used here as an
input parameter. %
\begin{figure}
\begin{centering}
\includegraphics{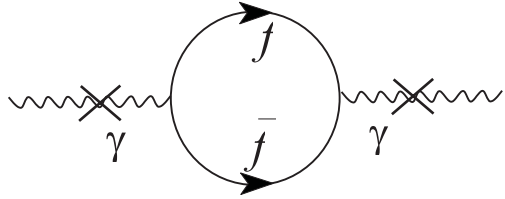} 
\par\end{centering}

\caption{Truncated self-energy graph in hadronic vacuum polarization\label{hdrvp}}

\end{figure}

Hadronic vacuum polarization $\widehat{\Pi}_{had}^{\gamma}\left(s\right)$
is related to the truncated $\left\{ \gamma-\gamma\right\} $ renormalized
self-energy (See Fig.(\ref{hdrvp})) by the following expression:
\begin{equation}
\widehat{\sum}_{ferm}^{\gamma}\left(s\right)=s\widehat{\Pi}_{had}^{\gamma}\left(s\right)+i\, Im\left(\widehat{\sum}_{ferm}^{\gamma}\left(s\right)\right),\label{e1.33}\end{equation}
 which can be easily evaluated by employing the free quark approximation.
An updated value of the dispersion integral, along with a logarithmic
parametrization, can be taken from \citep{Burkhardt2001}. A new reported
value coming from the light quark contribution at $s=m_{Z}^{2}$ is
$\bigtriangleup\alpha_{hadr}^{\gamma(5)}\left(m_{Z}^{2}\right)=-0.02761$.
This value can be reproduced by Eq.(\ref{e1.33}) using the following
masses of the light quarks: $m_{u}=m_{d}=53\, MeV$ (corresponds to
$\bigtriangleup\alpha_{hadr}^{\gamma(5)}\left(m_{Z}^{2}\right)=-0.027609$).
Clearly, the values of the light quark masses at low-Q scattering
processes should be adjusted by using this approach, but with $\bigtriangleup\alpha_{hadr}^{\gamma(5)}\left(s\right)$
calculated in the region of $\sqrt{s}<4.0$~GeV. The simple logarithmic
parametrization can be used here to extract quark masses at low momentum
transfer: \begin{equation}
\bigtriangleup\alpha_{hadr}^{\gamma(5)}\left(s\right)=A+B\ln\left(1+C\cdot s\right),\label{e1.34}\end{equation}
 with $A,B$ and $C$ parameters taken from \citep{Burkhardt2001}.
For low-momentum transfer experiments, the c.m.s. energy is $\sqrt{s}<4.0$
$GeV$, which gives, $m_{u}=m_{d}\simeq45\, MeV$.

\subsection{Vertex Corrections Graphs}

The vertex correction contributions can be split into two classes.
In the first class, where the electron vertex is at the one-loop level,
the amplitude is calculated according to \citep{BAB2002}. As in the
case of the self-energy graphs, the hadronic vertex does not belong
to the loop integrals, and therefore the PV amplitude is constructed
according to the SM Feynman rules employed in the \emph{FeynArts}
package. Generally, the electron vertex corrections have an infrared
divergence at $q\rightarrow0$ and are treated by the bremsstrahlung
contribution considered in \citep{ThesisAA}.

\begin{figure}
\begin{centering}
\includegraphics[scale=0.8]{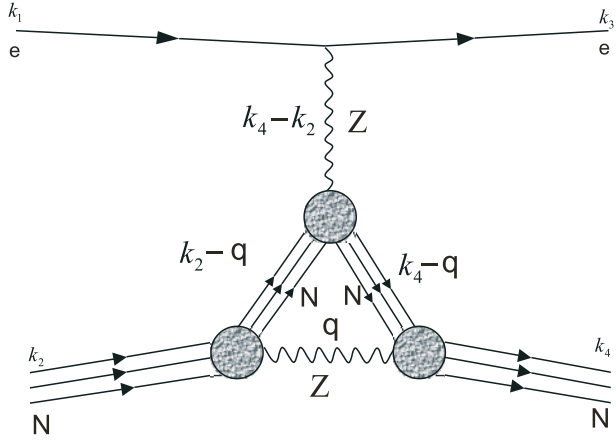} 
\par\end{centering}

\caption{Hadronic vertex correction graph.\label{vr.cr.Z.ex}}

\end{figure}

The second class of the triangle graphs are hadronic vertex corrections.
In this case, to evaluate the amplitude automatically using the packages
\emph{FeynArts} and \emph{FormCalc}, we construct an expansion similar
to that given for the box diagrams. To work out the set of rules for
the triangle topology, it is sufficient to consider the example in
Fig.(\ref{vr.cr.Z.ex}).%

\begin{figure}
\begin{centering}
\includegraphics[scale=0.3]{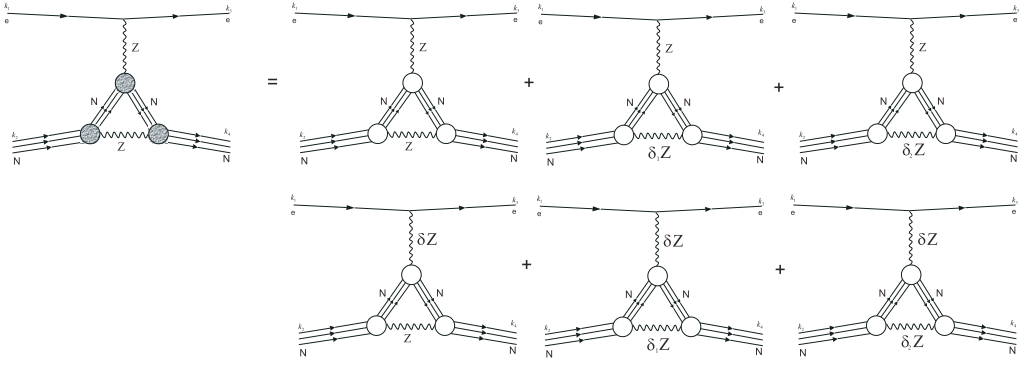} 
\par\end{centering}

\caption{Hadronic vertex expansion in the terms of $Z,\,\delta_{1}Z$ and $\delta_{2}Z$
partners.\label{hdr.vert.exp}}

\end{figure}

For the graph in Fig.(\ref{vr.cr.Z.ex}), the amplitude denominator
has the structure \begin{eqnarray}
\frac{1}{D_{1}D_{2}D_{3}D_{4}} & = & \frac{1}{\left(k_{4}-q\right)^{2}-m_{N}^{2}}\cdot\frac{1}{\left(k_{2}-q\right)^{2}-m_{N}^{2}}\cdot\frac{1}{\left(k_{4}-k_{2}\right)^{2}-m_{Z}^{2}}\cdot\nonumber \\
\label{e1.35a}\\ &  & \ \ \ \ \ \ \frac{1}{q^{2}-m_{Z}^{2}}\cdot\ \frac{\Lambda^{2}}{\left(k_{4}-k_{2}\right)^{2}-\Lambda^{2}}\cdot\frac{\Lambda^{4}}{\left(q^{2}-\Lambda^{2}\right)^{2}},\nonumber \end{eqnarray}
 which can be easily expanded into \begin{eqnarray}
\frac{1}{D_{1}D_{2}D_{3}D_{4}} & = & \frac{1}{\left(k_{4}-q\right)^{2}-m_{N}^{2}}\frac{1}{\left(k_{2}-q\right)^{2}-m_{N}^{2}}\cdot\nonumber \\
\nonumber \\ &  & \ \ \ \ \ \ \ \left(\frac{B^{Z-N}}{\left(k_{4}-k_{2}\right)^{2}-m_{Z}^{2}}+\frac{B^{\delta Z-N}}{\left(k_{4}-k_{2}\right)^{2}-\Lambda^{2}}\right)\cdot\nonumber \\
\label{e1.35}\\ &  & \ \ \ \ \ \ \ \lim_{\left\{ \Lambda_{1},\Lambda_{2}\right\} \rightarrow\Lambda}\ \left(\frac{\left(C^{Z-N}\right)^{2}}{q^{2}-m_{Z}^{2}}+\frac{\left(C^{\delta_{1}Z-N}\right)^{2}}{q^{2}-\Lambda_{1}^{2}}+\frac{\left(C^{\delta_{2}Z-N}\right)^{2}}{q^{2}-\Lambda_{2}^{2}}\right).\nonumber \end{eqnarray}
 Here, the coefficients $B^{Z-N}$ and $B^{\delta Z-N}$ are equal
to $\pm\sqrt{\frac{\Lambda^{2}}{\left(\Lambda^{2}-m_{Z}^{2}\right)}}$
respectively. $C^{Z-N},C^{\delta_{1}Z-\Lambda_{1}}$, and $C^{\delta_{2}Z-\Lambda_{2}}$
can be calculated using the following formulae: \begin{eqnarray}
\left(C^{Z-N}\right)^{2} & = & \frac{\Lambda^{4}}{\left(m_{Z}^{2}-\Lambda_{1}^{2}\right)\left(m_{Z}^{2}-\Lambda_{2}^{2}\right)},\nonumber \\
\nonumber \\\left(C^{\delta_{1}Z-N}\right)^{2} & = & -\frac{\Lambda^{4}}{m_{Z}^{2}-\Lambda_{1}^{2}}\frac{1}{\Lambda_{1}^{2}-\Lambda_{2}^{2}},\nonumber \\
\label{e1.36}\\\left(C^{\delta_{2}Z-N}\right)^{2} & = & -\frac{\Lambda^{4}}{m_{Z}^{2}-\Lambda_{1}^{2}}\left(\frac{1}{m_{Z}^{2}-\Lambda_{2}^{2}}-\frac{1}{\Lambda_{1}^{2}-\Lambda_{2}^{2}}\right).\nonumber \end{eqnarray}
 The expansion of the amplitude denominator in Eq.(\ref{e1.35}) has
a simple graphical representation (see Fig.(\ref{hdr.vert.exp}))
and suggests, in this particular case of the triangle topology, that
a set of virtual particles $\delta_{1}Z$ and $\delta_{2}Z$ should
be introduced in the NLO hadronic vertex corrections.%

Although ultraviolet divergences are absent in the hadronic vertex
corrections due to the additional terms proportional to $\frac{\Lambda^{2}}{q^{2}-\Lambda^{2}}$
in the coupling, it is still necessary to compute the wave function
renormalization with some details given in the appendix. As well as
in the case of electron vertex corrections, the nucleon vertex will
have an infrared divergence at the pole $q\rightarrow0$ (for the
proton). A detailed discussion of the treatment of this type of divergence
is given in \citep{ThesisAA}.

\section{Numerical Results and Conclusions}

The application of the methods described above for $\left\{ e-N\right\} $
scattering can be found in calculations of the weak charges of the
nuclei. Consider the parity-violating Hamiltonian in Eq.(\ref{i1.4}).
Here, for a heavy nucleus we have a coherent effect for $V\left(N\right)\otimes A\left(e\right)$:
\begin{equation}
\left(\overline{u}_{N}\gamma^{\mu}u_{N}\right)\rightarrow\rho_{nuc}\left(r\right)\delta_{\mu,0}.\label{d1.1}\end{equation}
 The contribution coming from $V\left(e\right)\otimes A\left(N\right)$
is small as it depends on unpaired valence nucleons. The latter determines
the Hamiltonian for the electron parity-violating interaction with
the nucleus in the following form: \begin{equation}
H\left(r\right)=\frac{G_{\mu}}{2\sqrt{2}}Q_{weak}\gamma_{5}\rho_{nuc}\left(r\right).\label{d1.2}\end{equation}
 A relation between the weak charge $Q_{weak}$ and form factors $\left\{ C_{1p},C_{1n}\right\} $
is straightforward: \begin{eqnarray}
Q_{weak}^{p} & = & 2C_{1p}\left(Q^{2}\rightarrow0\, GeV^{2}\right),\nonumber \\
\label{d1.3}\\Q_{weak}^{n} & = & 2C_{1n}\left(Q^{2}\rightarrow0\, GeV^{2}\right).\nonumber \end{eqnarray}
 If we take into account only the leading order of the interaction,
the weak charge of the proton and neutron have the simple definitions:
\begin{eqnarray}
Q_{weak}^{p\left(0\right)} & = & 1-4s_{w}^{2},\nonumber \\
\label{d1.4}\\Q_{weak}^{n\left(0\right)} & = & -1,\nonumber \end{eqnarray}
 and for the nucleus \begin{equation}
Q_{weak}=Z\cdot Q_{weak}^{p}+N\cdot Q_{weak}^{n},\label{d1.5}\end{equation}
 where $Q_{weak}^{p},$ $Q_{weak}^{n}$ are the weak charges of the
proton and neutron including NLO corrections.

\begin{figure}
\begin{centering}
\includegraphics[scale=0.57]{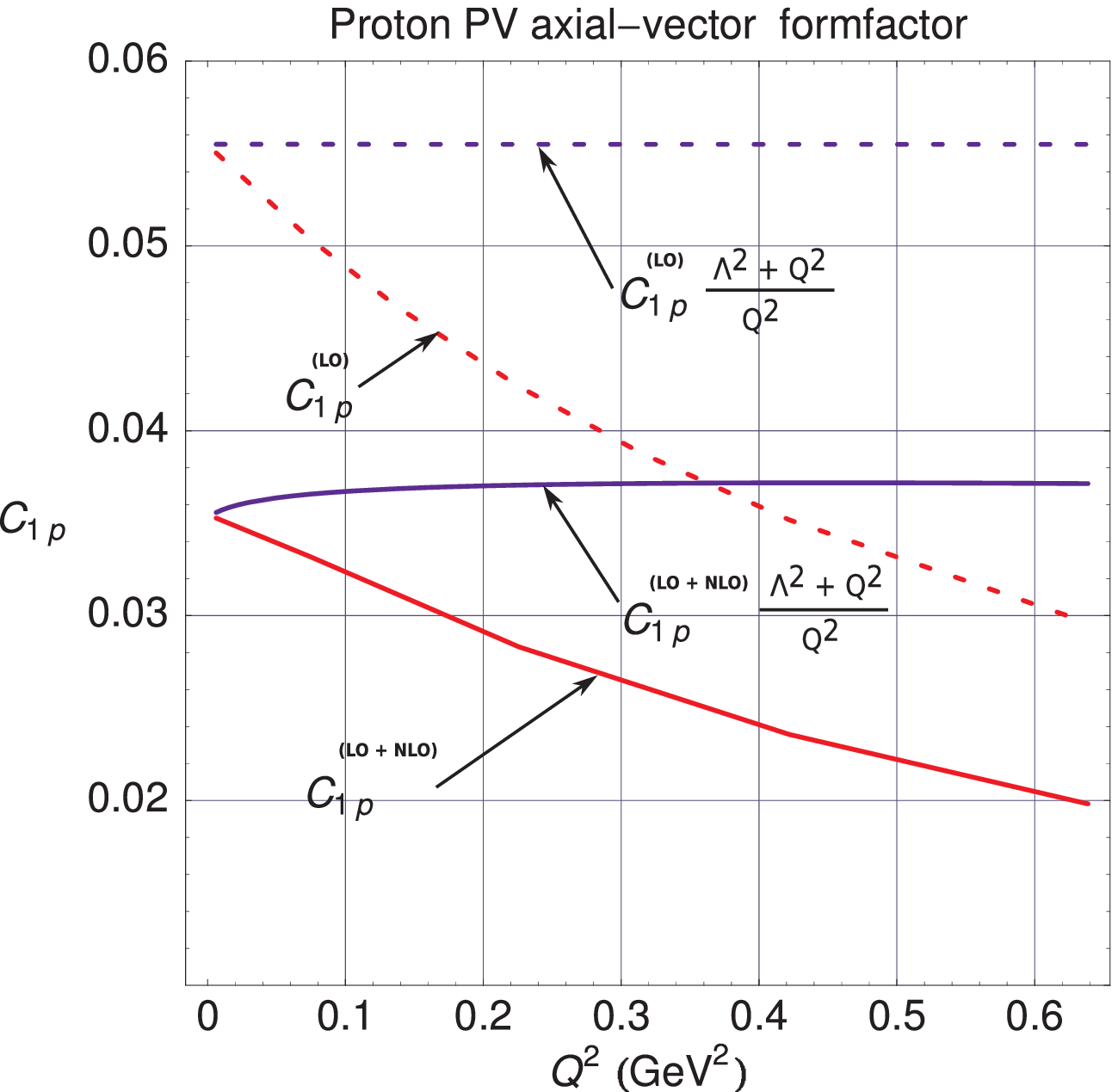} ~~~~~~\includegraphics[scale=0.54]{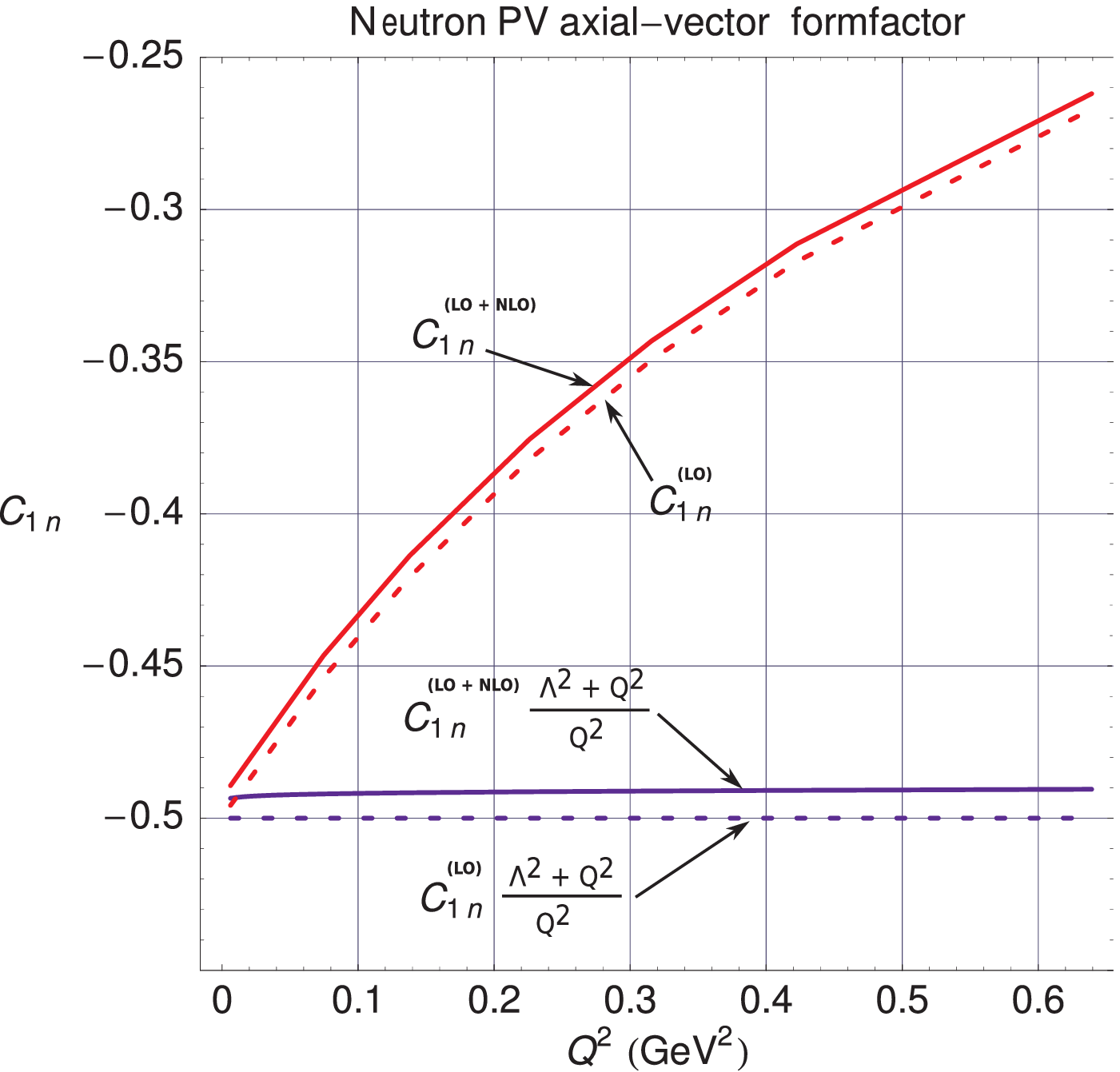} 
\par\end{centering}

\caption{Momentum transfer dependence of the form factors $\left\{ C_{1p},\, C_{1n}\right\} $
and $\left\{ C_{1p},\, C_{1n}\right\} \cdot\frac{\Lambda^{2}+Q^{2}}{Q^{2}}$.
The dotted line represents parity-violating form factor at the tree-level
(LO) only, and the solid line shows the result with NLO contribution
included (LO+NLO). \label{c1np.1}}

\end{figure}

As one can see from Fig.(\ref{c1np.1}), the input of the NLO corrections
is much more significant for the proton than for the neutron. This
makes experiments involving neutrons more interesting, as they allow
for the cleaner weak charge extraction.%

Also, form factors $C_{1p}$ and $C_{1n}$ normalized by the monopole
term $\frac{\Lambda^{2}}{\Lambda^{2}+Q^{2}}$ (see Fig.(\ref{c1np.1}))
do not exhibit strong momentum transfer dependence and hence are model
independent. This justifies to some extent our choice of the monopole
form factor approximation.

Extrapolation of the $C_{1p}$ and $C_{1n}$ to zero momentum transfer
point (see Fig.(\ref{c1np.1})) gives: $C_{1p}=0.0360\pm0.0005$ and
$C_{1n}=-0.4950\pm0.0020$. The most important of these numbers is
the result for $C_{1p}.$ It has been evaluated most recently in \citep{EKM2003},
leading to $C_{1p}=0.0358\pm0.0003$.

The available results from atomic parity-violating experiments for
the weak charges of $Cs_{55}^{133},\, Tl_{81}^{205}$ and $Bi_{83}^{209}$
(\citep{L2003} and references within, and \citep{BZ78}) can be used
as an experimental test of the theoretical predictions: \begin{eqnarray}
Cs_{55}^{133}(\exp) & = & -72.65\pm0.28\pm0.34,\nonumber \\
\label{d1.7}\\Tl_{81}^{205}(\exp) & = & -114.8\pm1.2\pm3.4,\nonumber \\
\nonumber \\Bi_{83}^{209}(\exp) & = & -140\pm40.\nonumber \end{eqnarray}
 Here, the errors are statistical, systematic and coming from an uncertainty
of the atomic-physics theory, respectively. For example, in the case
of $Cs_{55}^{133},$ one should observe $7s\;(excited)\rightarrow6s\;(ground)$
parity-violating electric dipole transitions in order to extract the
weak charge of $Cs_{55}^{133}.$ This requires an accurate knowledge
of the atomic wave functions. Some of the analysis of the issues involved,
with an extensive reference list, is given in \citep{PDG} and \citep{EKM2003}.
Using Eq.(\ref{d1.3}) and Eq.(\ref{d1.5}) we compute the following
results for the corresponding nuclear weak charges: \begin{eqnarray}
Cs_{55}^{133}(\mathrm{theory}) & = & -73.26\pm0.26,\nonumber \\
\label{d1.7a}\\Tl_{81}^{205}(\mathrm{theory}) & = & -114.9\pm0.4,\nonumber \\
\nonumber \\Bi_{83}^{209}(\mathrm{theory}) & = & -118.8\pm0.4,\nonumber \end{eqnarray}
 which clearly agree with the experiment. The theoretical uncertainty
is estimated quite generously and comes mostly from the numerical
integration and the extrapolation to the zero momentum transfer point.
Although the NLO corrections contribute only $\sim1.5\%$ to the results
in Eq.(\ref{d1.7a}), their value is still about four times larger
than our theoretical uncertainty. Thus, the more precise is the experiment
(like $Cs_{55}^{133}$), the more important it becomes to evaluate
most carefully the full set of the applicable NLO corrections. Although
our model predictions for the nuclear weak charges are in good agreement
with the available experimental results, to allow more definitive
conclusions about the validity of the proposed computational model,
the experimental errors for precision measurements of the weak charge
of the nucleon will have to be reduced.

Any significant deviation of the weak charge of the proton from the
Standard Model prediction at low $Q^{2}$ would be a signal of new
physics. The proton's weak charge $Q_{weak}^{p}$ is a well-defined
experimental observable. At $Q^{2}\rightarrow0$ the asymmetry can
be parametrized as

\[
\mathcal{A}=\left[\frac{-G_{F}Q^{2}}{4\sqrt{2}\pi\alpha}\right][Q^{2}Q_{weak}^{p}+Q^{4}B(Q^{2})],\]
 where $B(Q^{2})$ is a function of Sachs electromagnetic form factors
$G_{E,M}^{\gamma}$ related to the Dirac and Pauli form factors by
the following expression: \begin{equation}
G_{E}^{\gamma}=F_{1}-\frac{Q^{2}}{4m_{N}^{2}}F_{2},\,\,\, G_{M}^{\gamma}=F_{1}+F_{2}.\end{equation}

The measurement of $Q_{weak}^{p}$ will be done by the $Q_{weak}$
collaboration \citep{Qweak}, and may have extremely interesting physical
implications. See, for example, \citep{EKM2003}. It will allow one
to determine the proton's weak charge with $\simeq$4\% combined statistical
and systematic errors, which leads to $9\sigma$ for the running of
the weak mixing angle. The Standard Model evolution predicts a shift
of $\Delta sin^{2}\theta_{W}=+0.007$ at low $Q^{2}$ with respect
to the $Z^{0}$ pole best fit value of $0.23113\pm0.00015.$ The weak
mixing angle at the energy scale close to the $Z^{0}$ pole was measured
very precisely, but a precision experimental study of the evolution
of $sin^{2}\theta_{W}$ to lower energies still has to be carried
out. The asymmetry measurements proposed for $Q_{weak}$ experiment
will go as low as $Q^{2}=0.03\, GeV^{2}$, making it a very competitive
experiment.

Using the Dirac and Pauli form factors, we can now compute the extensive
set of one-loop hadronic electroweak NLO corrections along with the
weak charges of the proton and neutron. Using a monopole approximation
for the form factors, we modified general electroweak couplings by
inserting the appropriate form factors into the vertices. The monopole
structure of the form factors allows us to substitute one Feynman
diagram with {}``structured'' nucleon with a set of diagrams involving
only point-like nucleon. This expansion can be visualized by adding
fictitious additional vector bosons of mass $\Lambda$. For renormalization,
we choose the on-shell renormalization scheme.

In conclusion, it is evident that computational symbolic packages
such as \emph{FeynArts} and \emph{FormCalc} can be efficiently adapted
for the theoretical evaluation of NLO effects in electron-nucleon
scattering. Refs \citep{ABB2005_a} and \citep{ABB2006} give some
additional details. A newly proposed measurement of the electron weak
charge in parity-violating Moller scattering at 12 GeV at JLab could
be a very interesting challenge, for example.

\begin{acknowledgments}
The authors thank Malcolm Butler of Saint Mary's University for his
useful comments. We are also grateful to Shelley Page of University
of Manitoba for some of her explanations on the $Q_{weak}$ experiment.
This work has been supported by NSERC (Canada). 
\end{acknowledgments}

\section{Appendix}

\subsection{Some analytical details on couplings}

\begin{figure}[H]
\begin{centering}
\includegraphics[scale=0.4]{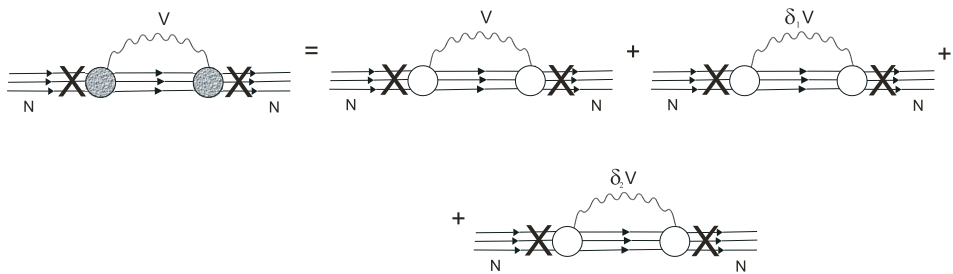}
\par\end{centering}

\caption{Hadronic wave function renormalization expansion.\label{hdr.nucl.ren.cons}}

\end{figure}

To introduce couplings between vector bosons and nucleon into the
model files of \emph{FeynArts, }instead of vector and axial-vector
representation, we employ use of chirality projectors. That can be
achieved if we compare Eq.(\ref{e1.1}), Eq.(\ref{e1.2}) with Eq.(\ref{e1.6}),
combined with Eq.(\ref{e1.7}) and Eq.(\ref{e1.8}), then it is possible
to write \begin{eqnarray}
\Gamma_{\mu}^{Z-N}\left(q\right)=ie[\frac{g_{R}^{Z-N}\left(q\right)+g_{L}^{Z-N}\left(q\right)}{2}\gamma_{\mu}+\frac{g_{R}^{Z-N}\left(q\right)-g_{L}^{Z-N}\left(q\right)}{2}\gamma_{\mu}\gamma_{5}+\nonumber \\
\nonumber \\+\frac{i}{2m_{N}}\sigma_{\mu\alpha}q^{\alpha}f_{2}\left(q\right)] & =\nonumber \\
\label{e1.11}\\=ie\left[g_{L}^{Z-N}\left(q\right)\gamma_{\mu}\varpi_{-}+g_{R}^{Z-N}\left(q\right)\gamma_{\mu}\varpi_{+}+\frac{i}{2m_{N}}\sigma_{\mu\alpha}q^{\alpha}f_{2}\left(q\right)\right]\nonumber \\
\nonumber \\\Gamma_{\mu}^{\gamma-N}\left(q\right)=ie\left[\frac{g_{L}^{\gamma-N}\left(q\right)+g_{R}^{\gamma-N}\left(q\right)}{2}\gamma_{\mu}+\frac{i}{2m_{N}}\sigma_{\mu\alpha}q^{\alpha}F_{2}\left(q\right)\right]=\nonumber \\
\label{e1.12}\\=ie\left[g_{L}^{\gamma-N}\left(q\right)\gamma_{\mu}\varpi_{-}+g_{R}^{\gamma-N}\left(q\right)\gamma_{\mu}\varpi_{+}+\frac{i}{2m_{N}}\sigma_{\mu\alpha}q^{\alpha}F_{2}\left(q\right)\right], &  & \nonumber \end{eqnarray}
 where \begin{eqnarray}
g_{L,R}^{Z-N}\left(q\right) & = & \frac{1}{4c_{w}s_{w}}\left(F_{1}^{V\left(N\right)}\left(0\right)-4s^{2}F_{1}\left(0\right)\pm g_{A}\left(0\right)\right)\frac{\Lambda^{2}}{\Lambda^{2}-q^{2}},\nonumber \\
\label{e1.13}\\g_{L,R}^{\gamma-N}\left(q\right) & = & g^{\gamma-N}\left(q\right)=F_{1}\left(0\right)\frac{\Lambda^{2}}{\Lambda^{2}-q^{2}}.\nonumber \end{eqnarray}
 In Eqs.(\ref{e1.11}) and (\ref{e1.12}), we have adopted the general
structure of the coupling from Eq.(\ref{e1.1}), with coupling strengths
$g_{L,R}^{Z-N}\left(q\right)$ and $g_{L,R}^{\gamma-N}\left(q\right)$
given by Eq.(\ref{e1.13}). Vector part of the coupling $\sigma_{\mu\alpha}q^{\alpha}$
also can be replaced by the chirality projectors $\omega_{\pm}$ in
the following way\begin{eqnarray}
\sigma_{\mu\alpha}q^{\alpha} & = & \frac{i}{2}\left(\left[\gamma_{\mu},\not q\right]\varpi_{-}+\left[\gamma_{\mu},\not q\right]\varpi_{+}\right).\label{e1.13b}\end{eqnarray}
 Moreover, to adopt \emph{FeynArts }representation of the coupling
through the product of generic and class type of the couplings, we
introduce the following matrix representation of Eq.(\ref{e1.11})
and Eq.(\ref{e1.12}): \begin{equation}
\Gamma\left(N,N,V_{\mu}\right)=\left(\begin{tabular}{llll}
 $\gamma_{\mu}\varpi_{-}$,  &  $\gamma_{\mu}\varpi_{+}$,  &  $\left[\gamma_{\mu},\not q\right]\varpi_{-}$,  &  $\left[\gamma_{\mu},\not q\right]\varpi_{+}$\end{tabular}\right)\overrightarrow{G}_{NNV}\left(q\right),\label{i1.1}\end{equation}
 with $\overrightarrow{G}_{NNV}$ defines coupling between classes
of the vector bosons and nucleons and expressed as a $2\times4$ matrix
\begin{equation}
\overrightarrow{G}_{NNV}=ie\left(\begin{array}{cc}
g_{L}^{V-N}\left(q\right) & G_{1L}^{V-N}\\
g_{R}^{V-N}\left(q\right) & G_{1R}^{V-N}\\
-\frac{1}{4m_{N}}F_{2}^{V-N}\left(q\right) & G_{2L}^{V-N}\\
-\frac{1}{4m_{N}}F_{2}^{V-N}\left(q\right) & G_{2R}^{V-N}\end{array}\right).\label{i1.2}\end{equation}
 The second column of $\overrightarrow{G}_{NNV}$ represents counterterms
of the first order. The Pauli form factor $F_{2}^{V-N}$ in Eq.(\ref{i1.2})
has the following structure: \begin{eqnarray}
F_{2}^{Z-N}\left(q\right) & = & f_{2}\left(q\right),\nonumber \\
\label{i1.3}\\F_{2}^{\gamma-N}\left(q\right) & = & F_{2}\left(q\right).\nonumber \end{eqnarray}
 The coupling defined in Eq.(\ref{i1.2}) has a counterterm part at
the one-loop level represented by the matrix, which has following
structure: \begin{equation}
\left(\begin{array}{c}
G_{1L}^{V-N}\\
G_{1R}^{V-N}\\
G_{2L}^{V-N}\\
G_{2R}^{V-N}\end{array}\right)=\left(\begin{array}{c}
g_{L}^{V-N}Re[\delta\, f_{L}]\\
g_{R}^{V-N}Re[\delta\, f_{R}]\\
-\frac{1}{4m_{N}}F_{2}^{V-N}\left(q\right)Re[\delta\, f_{L}]\\
-\frac{1}{4m_{N}}F_{2}^{V-N}\left(q\right)Re[\delta\, f_{R}]\end{array}\right),\label{e1.37}\end{equation}
 where the hadronic field renormalization constants $\delta\, f_{L,R}$
are computed using the expansion on Fig.(\ref{hdr.nucl.ren.cons}).
More details are given in Ref.\citep{ThesisAA}.%

\end{document}